# *VOCAL SIGNAL DIGITAL PROCESSING. INSTRUMENT FOR ANALOG TO DIGITAL CONVERSION STUDY.*

**Ovidiu-Andrei SCHIPOR[1], Felicia-Florentina GÎZĂ [2]**

*Universitatea "Ştefan cel Mare" Suceava*
*Str. Universităţii nr.9, 720225, Suceava, Romania*
*[1]schipor@eed.usv.ro, [2]felicia@eed.usv.rom*

***Abstract*** *The goal of this article is to present interactive didactic software for analog to digital conversion using PCM method. After a short introduction regarding vocal signal processing we present some method for analog to digital conversion.*
*The didactic software is an applet that can be direct accessed by any interested person.*

## 1.Prelucrarea numerică a semnalelor

Procesarea numerică de semnal (PNS) se referă la o varietate de tehnici de îmbunătăţire a acurateţei şi siguranţei comunicaţiilor digitale. Teoria care stă la baza prelucrării numerice de semnal este suficient de complexă, însă la baza acestei tehnici de procesare se găseşte stabilirea şi standardizarea nivelelor şi a stărilor unui semnal digital.
Semnalele analogice reprezintă o mărime care se modifică in mod continuu (de exemplu intensitatea sonoră a unui ton). În schimb, semnalele digitale pot prelua numai anumite valori (discrete), de exemplu valorile 0 şi 1 respectiv „sub tensiune" sau „fără tensiune".
Toate circuitele de comunicaţie conţin zgomot. Acest lucru este valabil indiferent dacă semnalele sunt analogice sau digitale şi indiferent de tipul de informaţie transmis. Zgomotul este principala sursă de neplăceri pentru inginerii de comunicaţie care încearcă mereu să găsească noi metode de îmbunătăţire a raportului semnal/zgomot (S/N) în sistemele de comunicaţie. Metodele tradiţionale de optimizare a raportului semnal/zgomot includ creşterea puterii semnalului transmis şi creşterea sensibilităţii receptorului de semnal.
Principalul avantaj al utilizării semnalelor digitalizate este că orice prelucrare ulterioară a acestora este în principiu lipsită de pierderi de informaţie (numerele nu sunt afectate de zgomot iar precizia calculelor poate fi matematic controlată).
Un sistem de prelucrarea numerică a semnalelor îndeplineşte în esenţă un ansamblu de operaţii şi anume:

- conversia semnalului analogic în semnal numeric
- prelucrarea semnalului numeric obţinut
- conversia semnalului numeric prelucrat în semnal analogic.

Dacă un semnal de intrare este analog, acesta este mai întâi convertit într-o formă digitală de un convertor analog-digital (CAD). Semnalul rezultat are două sau mai multe nivele. Ideal, aceste nivele sunt cunoscute exact, reprezentând curenţi sau tensiuni. Totuşi, deoarece semnalul de intrare conţine zgomot, nivelele nu sunt întotdeauna egale cu valorile standard. Circuitele de prelucrare a semnalului ajustează aceste nivele astfel încât să reprezinte valorile corecte. De fapt acestea elimină zgomotul.

## 2.Tehnici de conversie analog-numerică

Conversia analogică digitală este un proces electronic în care un semnal variabil continuu în timp şi amplitudine (analog) este transformat (cu o eroare controlată), într-un semnal digital.
Intrarea unui convertor analogic digital (CAD) constă într-un nivel de tensiune care variază într-un interval infinit de valori. Exemple sunt: formele de undă sinusoidale, formele de undă care reprezintă vorbirea umană, semnalele de la o cameră video etc.
Semnalele digitale sunt propagate mult mai eficient decât semnalele analogice, în mare parte datorită faptului că impulsurile digitale, care sunt foarte bine definite şi ordonate, sunt mult mai uşor de deosebit de zgomot, care este haotic. Acesta este avantajul principal al modurilor de comunicare digitală.

**2.1.Modulaţia în cod a impulsurilor (PCM)**

Marele progres în fabricarea componentelor electronice, în special a circuitelor înalt integrate şi cel mai înalt integrate, îl reprezintă utilizarea frecventă a tehnicii numită Pulse Code Modulation (PCM – modulaţia impulsurilor în cod) în tehnica informaţiilor şi în electronică, precum şi în electronica de divertisment, ca de exemplu la video-disc şi la discul digital (compact-disc). PCM este unul din cele mai vechi şi conceptual cele mai simple procese de conversie de la analogic la digital utilizate în semnalele vocale şi video.
În cazul modulării în cod a impulsurilor (PCM), semnalele care se prelucrează (de exemplu tonuri sonore) nu se prezintă sub forma oscilaţiilor, ci sub forma numerelor binare formate din mai mulţi biţi care au fost obţinute ca rezultat al eşantionării şi cuantizării cursului oscilaţiilor.
Eşantionarea are loc de cele mai multe ori cu ajutorul unui aşa-numit *circuit sample-and-hold* (eşantionare-şi-menţinere) care înregistrează mărimile continue de intrare (de exemplu valorile tensiunii unei oscilaţii electrice) sub formă de semnale periodice de foarte scurtă durată. Semnalul format prin eşantionare este transformat apoi într-o mărime digitală de ieşire prin intermediul unităţilor de cuantizare şi codare.
Cu ajutorul modulării în cod a impulsurilor (PCM) este posibilă recunoaşterea, acoperirea sau corectarea transmiterii erorilor in cazul transmiterii semnalului. Valorile răspunzătoare de erori pot fi eliminate prin prelucrarea semnalului şi înlocuire prin media valorilor corecte învecinate.

**2.2.Modulaţia numerică diferenţială**

În cadrul modulaţiei numerice diferenţiale (DNUM), în locul informaţiei despre un anumit eşantion se transmite o informaţie despre diferenţa dintre acesta şi un eşantion determinat prin predicţie.
Cele mai cunoscute tehnici de modulaţie numerică diferenţială sunt:

- modulaţia diferenţială a impulsurilor în cod (DPCM) la care sunt generate doar diferenţele dintre amplitudini consecutive (în acest mod se diminuează rata de transfer necesară);
- modulaţia delta (ΔM) la care se generează un singur bit de diferenţă între amplitudini succesive (acest bit indică creşterea / descreşterea eşantionului curent faţă de eşantionul precedent).

**2.2.Modulaţia numerică adaptativă**

În cadrul modulaţiei numerice adaptive (ADPCM) se realizează corespondenţa de la eşantion la alfabet funcţie de istoria semnalului. În acest scop se constituie o stare $S_n$ a sistemului în intervalul nT, iar corespondenţa între valoarea eşantionului x(nT) şi valoarea discretă $y_k$ se va face ţinând cont şi de această stare. Acest procedeu de modulaţie este cel mai eficient.

**3.Conversia analog-numerică prin metoda PCM**

Prin conversia analog-numerică, semnalul continuu este eşantionat şi cuantizat, fiecare nivel obţinut fiind reprezentat printr-un cuvânt binar care se aplică la intrarea sistemului de prelucrare numerică. Această primă fază a prelucrării numerice este realizată cu convertoare specializate analog-numerice şi este deosebit de importantă, deoarece prin aproximările pe care le efectuează, contribuie la precizia de calcul şi la raportul semnal zgomot final.

### 3.1.Eşantionarea

Fie $x_a(t)$ un semnal analogic (continuu în timp) şi $\{t_n\}_{n\in Z}$ o mulţime numărabilă de valori reale distincte ordonate ($t_n < t_m$ dacă $n < m$). Eşantionarea este transformarea semnalului $x_a(t)$ în semnalul discret x[n] definit prin relaţia:

$$x[n] = x_a(t_n) \quad (1)$$

Eşantionarea uniformă este dată de relaţia:

$$x[n] = x_a(nT) \quad (2)$$

unde T>0 este perioada de eşantionare, iar $t_n = n*T$.

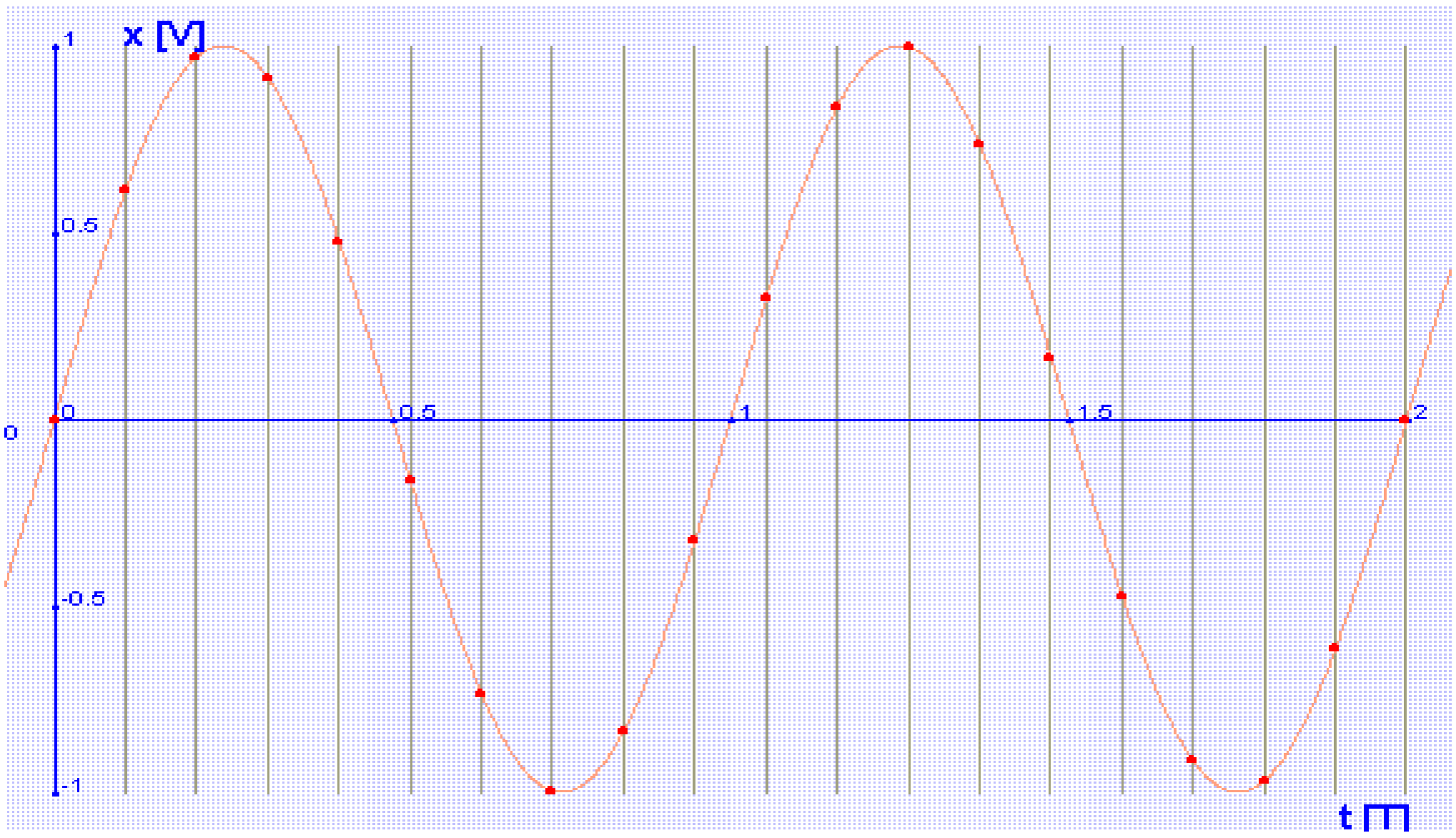

**Figura 1. Eşantionarea (semnal sinusoidal, doua perioade, 20 eşantioane)**

Prin eşantionare reţinem numai valorile continue în amplitudine şi discrete în timp. În mod ideal, eşantionarea nu are ca rezultat o pierdere de informaţie şi nici nu introduce distorsiuni în semnal dacă sunt respectate condiţiile teoremei eşantionării. Un semnal analogic poate fi refăcut din eşantioanele sale dacă a fost eşantionat la o frecvenţă de cel puţin două ori mai mare decât lărgimea de bandă a semnalului analogic (lărgimea de bandă = frecvenţa superioară – frecvenţa inferioară).
De exemplu, pentru muzică, al cărui domeniu de frecvenţe se situează între 20 si 20 000 Hz (limitele analizatorului auditiv uman), sunt necesare deci cel puţin 40 000 de eşantioane pe secundă. Pentru compact-disc, semnalul se eşantionează, de exemplu, cu 44 100 de eşantioane pe secundă si astfel sunt reţinute tot atât de multe valori pe secundă.
Vocea umană, poate fi redată optim prin sunete cu frecvenţe cuprinse între 100 şi 8.000 Hz.(limitele aparatului fonoarticulator). Acesta este motivul pentru care sistemele de telefonie au o gamă de frecvenţe de răspuns relativ îngustă, eliminând sunetele de înaltă frecvenţă. Drept rezultat, sunetul înregistrat de un sistem de recunoaştere a vorbirii poate fi eşantionat la o rată minimă de numai 8kHz, cu toate că 16kHz ar putea oferi rezultate mai bune, dacă sistemul dispune de suficientă putere de procesare şi de stocare de date.
La ieşirea dispozitivului de eşantionare se obţine o secvenţă de impulsuri. Amplitudinea fiecărui impuls este proporţională cu amplitudinea semnalului de intrare analogic în momentul eşantionării. Din acest motiv, acest pas se numeşte modulaţie de amplitudine a impulsurilor.

### 3.1.Cuantizarea

Cuantizarea reprezintă transformarea semnalului continuu în amplitudine şi discret în timp într-un semnal discret în timp şi în amplitudine. Este un proces ireversibil care transformă amplitudinile notate cu x[n] sau $x_a[nT]$ în valori $y_k$ dintr-un set finit de valori.
Fie D domeniul semnalului de intrare. Acesta este împărţit în L intervale:

$$I_k = \{x_k < x[n] <= x_{k+1}\},\ k = 1, 2 \ldots L \qquad (3)$$

Având în vedere acest lucru, se obţin nivelele de cuantizare notate cu $y_1$, $y_2$ ...$y_k$ astfel:

$$x_q[n] = Q(x[n]) = y[n] = y_k, \qquad (4)$$

pentru $x[n] \in I_k$. , iar $Q(x[n])$ reprezentând intervalul $I_k$ în care se găseşte $x[n]$.

Diferenţa dintre $x[n]$ şi $x_q[n]$ se numeşte eroare de cuantizare sau zgomot de cuantizare:

$$e_q = x[n] - x_q[n] \qquad (5)$$

Eroarea de cuantizare nu poate depăşi o jumătate din pasul de cuantizare:

$$-\frac{\Delta}{2} \leq e_q \leq \frac{\Delta}{2} \qquad (6)$$

unde $\Delta =$ pas de cuantizare.

În cazul în care eroarea de cuantizare depăşeşte limitele admise, trebuie mărit numărul de nivele de cuantizare.

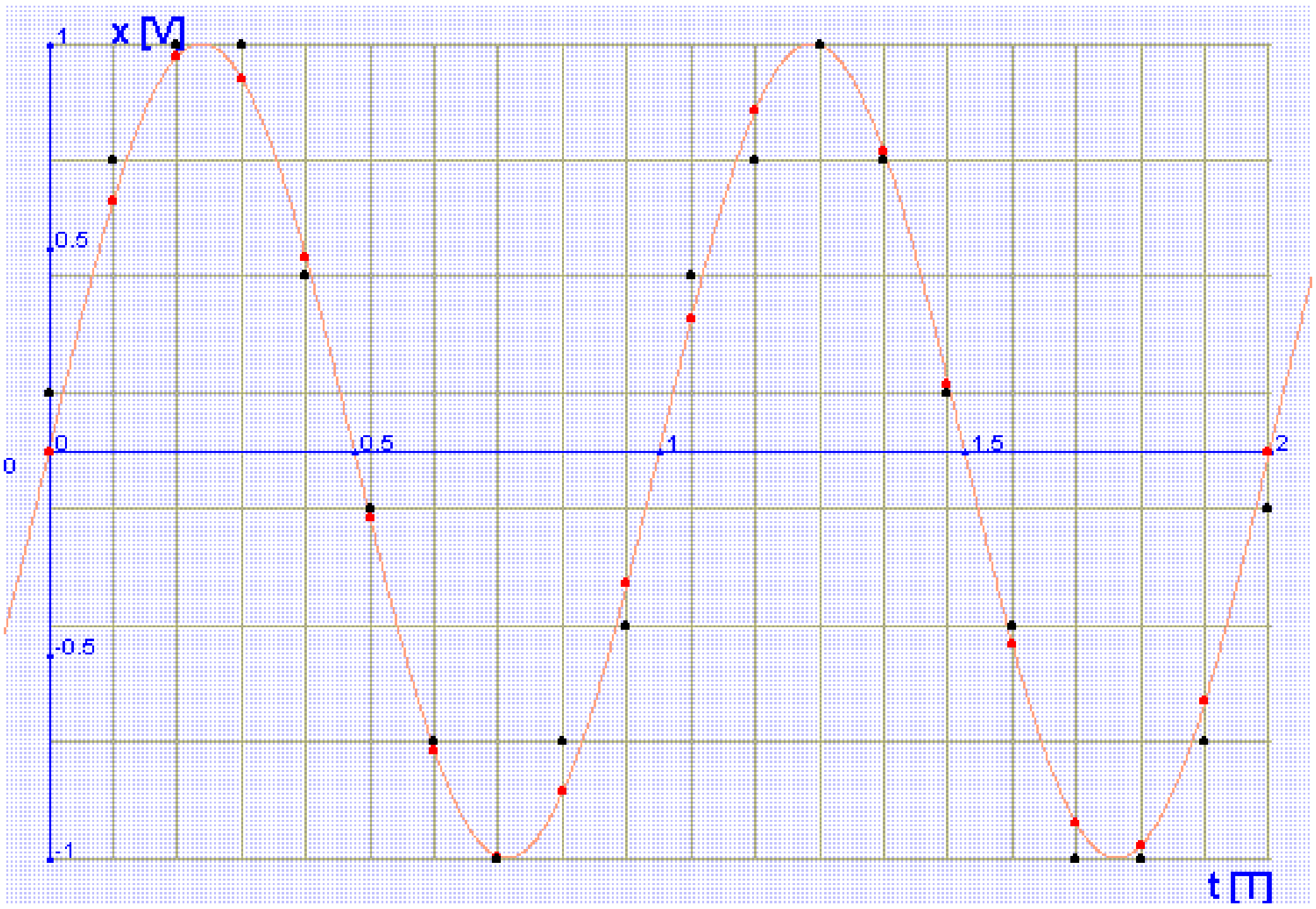

**Figura 2. Cuantizarea (8 nivele de cuantizare)**

### 3.3.Codarea

Codarea este procesul prin care fiecărei valori discrete $x_q[n]$ i se atribuie o secvenţă egală cu b biţi. Pentru codificarea celor k nivele de cuantizare posibile sunt necesari $\log_2 k$ biţi.

Apropierea necesară faţă de procesul oscilatoriu reclamă o gradare fină a valorilor măsurătorilor rezultate prin eşantionare, rezultând deci şiruri relativ lungi de biţi (lungimi de cuvinte). Astfel, printr-o lungime de cuvânt de 4 biţi, pot fi redate numai 7% din procesele oscilatorii.

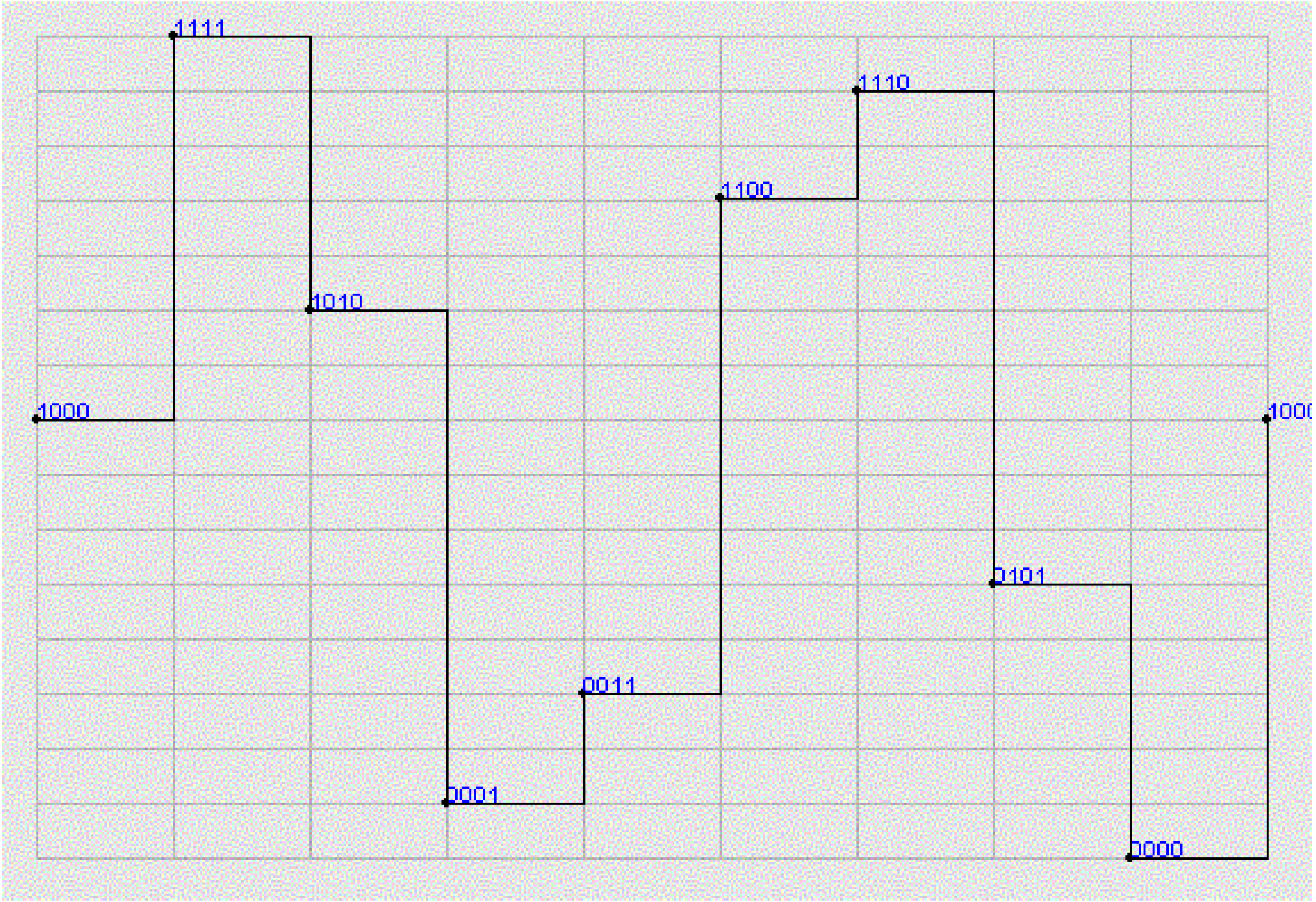

**Figura 3. Codificarea (în cazul folosirii a 16 nivele de cuantizare)**

Pentru un semnal cu frecvenţa de eşantionare de f Hz şi k nivele de cuantizare este necesară o rată de transfer:

$$RT = f * \log_2 k \text{ bps} \qquad (7)$$

De exemplu pentru transmiterea unui semnal eşantionat cu rata de eşantionare de 8 KHz şi codat pe 8 biţi (adică are 256 nivele de cuantizare) este necesară o rată de transfer de 64 Kbps. Se observă că o creştere a numărului de nivele de cuantizare necesită o rată de transfer mai mare.
Gama dinamică (diferenţa de volum dintre cel mai slab şi cel mai puternic sunet) pentru vocea umană este mult mai mică decât pentru muzica de înaltă calitate. În cele mai multe cazuri este nevoie de doar 8 biţi pe eşantion cu toate că rezultatele sunt mult mai bune în cazul folosirii a 16 biţi pe eşantion, care este şi numărul de biţi caracteristic CD-urilor audio.
Diferenţele dintre ratele de eşantionare şi numărul de biţi pot avea un impact major asupra cantităţii de date pe care calculatorul trebuie să o proceseze. O secundă de sunet digital la 8 kHz şi 8 biţi pe eşantion înseamnă doar 8.000 octeţi de date. Aceeaşi secundă de sunet digital la 16 kHz şi 16 biţi înseamnă de patru ori mai multe date: 32.000 octeţi. Standardul pentru CD-uri audio, de 44 kHz şi 16 biţi, înseamnă că o secundă de sunet necesită un spaţiu de stocare de 88.000 octeţi.
Pentru rezultate superioare, cuantizarea nu ar trebui efectuată uniform. Unele semnale sunt de amplitudini joase, iar altele sunt de amplitudini ridicate. În practică, cuantizarea este neuniformă, existând mai multe nivele de cuantizare pentru amplitudinile care predomină în cadrul semnalului.
Pentru o transmisie a semnalelor inteligibilă şi de o calitate acceptabilă a comunicaţiei se poate realiza reducerea vitezei de kbit/s cu algoritmi de codare şi de cuantizare vectorială. Scopul acestor algoritmi este de a transmite, memora şi sintetiza semnalul vocal de o calitate dată, utilizând mai puţini biţi. Această reducere este realizată eliminând redundanţa din semnalul vocal

**4. Instrument didactic pentru studiul conversiei analog-numerice prin metoda PCM**

Apletul a fost realizat utilizând limbajul Java, versiunea Sun SDK 1.4.
O parte din clasele utilizate au fost realizate de către autori în cadrul bursei de studiu ERASMUS-SOCRATES EUDIL-Lille Franta.
Interfaţa este realizată utilizând clasele din pachetul Abstract Window Toolkit.

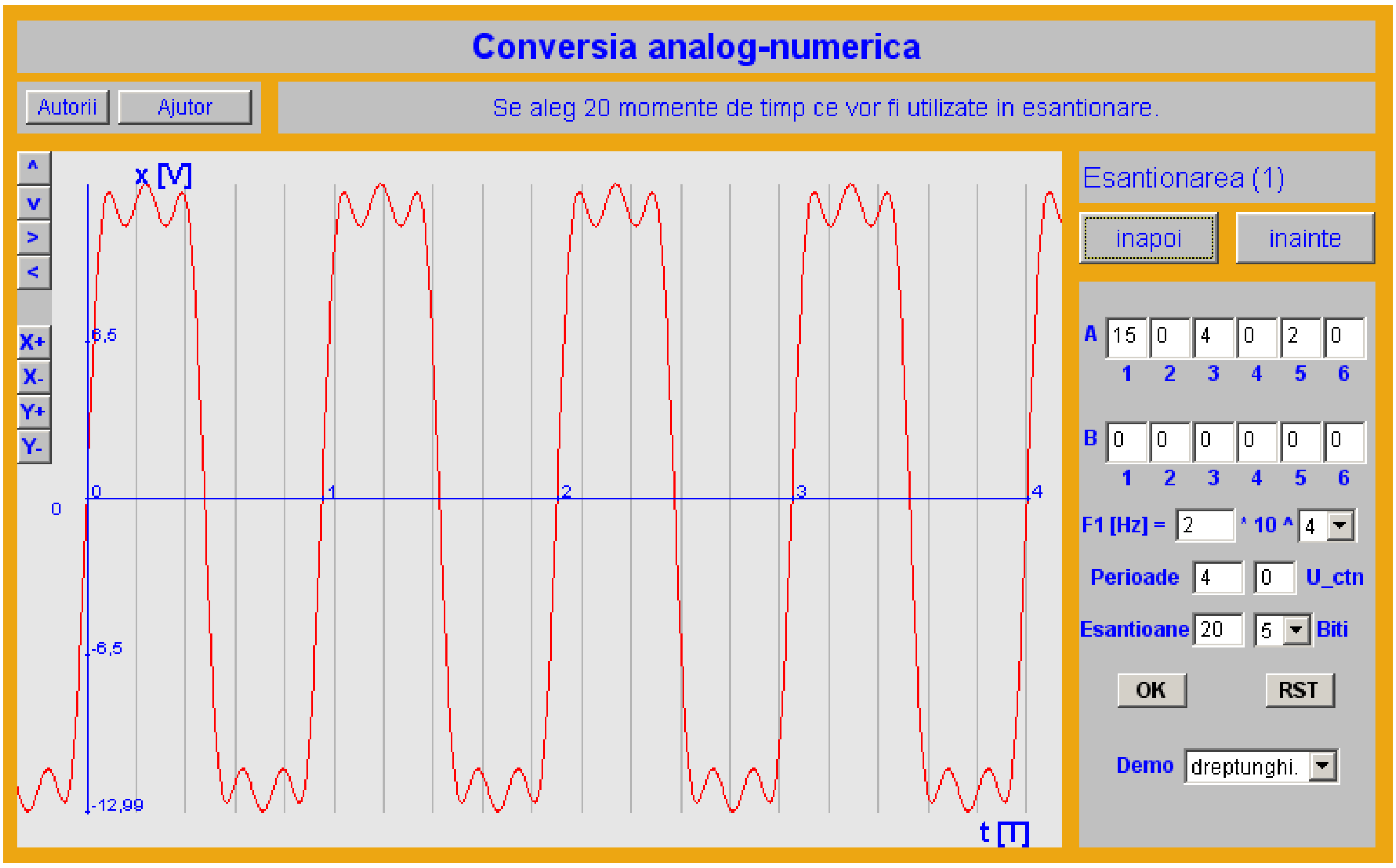

**Figura 4. Instrument didactic pentru studiul conversiei analog-numerice**

Apletul permite generarea unui semnal periodic nesinusoidal pe baza primelor 6 armonici. Ulterior, acest semnal este convertit numeric, pas cu pas, evidenţiindu-se aspectele esenţiale.
Apletul se utilizează în modul următor:

1. Se introduc coeficienţii A1…A6 (sin) şi B1…B6 (cos) ai primelor 6 armonici ale semnalului analogic.
2. Se introduce frecvenţa armonicii fundamentale (F1) în format mantisă - ordin de marime
3. Se introduce numărul de perioade (Perioade) ale armonicii fundamentale ce se doresc vizualizate.
4. Se introduce componenta constantă (U_ctn).
5. Se introduce numărul eşantioanelor de timp (Eşantioane).
6. Se introduce numărul de biţi utilizaţi pentru cuantizare (Biţi).
7. Se apasă butonul <OK> pentru a vizualiza semnalul analogic.
8. Se apasă butoanele <înainte> şi <înapoi> pentru vizualizarea succesivă a etapelor.

Butonul <RST> permite aducerea apletului în starea iniţială.
Butoane de ajutor: <Autorii> şi <Ajutor>.
Opt butoane permit o mai bună vizualizare a graficelor:
- ^ , v , > , < - pentru deplasarea graficului;
- X+ , X-, Y+ , Y- pentru zoom pozitiv şi negativ pe axa OX (a timpului) şi OY (a valorii semnalului).

Pentru o utilizare rapidă (fără introducere de date) se poate selecta un exemplu:

- sinusoidă – un semnal sinusoidal pe 2 perioade;
- triunghiular - un semnal triunghiular pe 2 perioade;
- dreptunghiular - un semnal dreptunghiular pe 4 perioade;
- o perioadă – un semnal oarecare pe o perioadă.

Semnalele triunghiular şi dreptunghiular au fost obţinute în limita preciziei oferite de 6 armonici.

**Referinţe bibliografice**

**cărţi**

[1] Dr. Kamilo Feher (1993) - *Comunicaţii digitale avansate*, Editura Tehnică, Bucureşti, Romania

[2] Constantin Ioan, Marghescu Ion (1995) - *Transmisiuni analogice şi digitale*, Editura Tehnică, Bucureşti, Romania

[3] Cornelia Marcuta, Mihai Creţu (2002) - *Măsurări electrice şi electronice*, Editura Tehnică-Info, Chişinău, Republica Moldova

**articole**

[4] Zoran Cvetkovic, Martin Vetterli (2004) - *Error rate characteristics of oversampled analog to digital conversion*, IEEE Transactions on Information Theory vol.44, nr. 5

[5] A.L. Wicks (2003) – *Current techniques of measurement, acquisition and processing test data*, 2003, Department of Mechanical Engineering Virginia Tech

**site-uri web**

[6] www.phys.ualberta.ca/~gingrich

[7] www.stanford.edu/ccrma/courses

[8] www.digital-recordings.com/publ/pubrec.html